\input{aipcheck}
\documentclass[
    ,final            
  ]
  {aipproc}

\layoutstyle{8x11double}

\usepackage{axodraw}
\begin{document}

\title{The Origin of Neutrino Masses and Physics Beyond the SM
\footnote{Plenary Talk given at NuFact09, Fermilab and IIT, Chicago, July 20-25, 2009.}}


\author{Pavel Fileviez P\'erez}{
  address={University of Wisconsin-Madison, 1150 University Ave, Madison WI 53706, USA}}

\begin{abstract}
We discuss the simplest models for the generation of neutrinos masses at tree level 
and one-loop level. The realization of the different seesaw mechanisms in the context 
of renormalizable $SU(5)$ and $SO(10)$ theories is reviewed. A new mechanism 
for the generation of neutrino masses at one-loop level is presented. We discuss the first 
realization of the Type III seesaw mechanism in the context of a renormalizable 
$SU(5)$ theory, called ``Adjoint SU(5)".  
\end{abstract}

\maketitle

\section{Introduction}
The existence of massive neutrinos is one of the main motivations for physics beyond 
the Standard Model. There are many possible theoretical frameworks where one could 
understand the origin of neutrino masses. In my opinion, the simplest way to classify these 
scenarios is using the $B-L$ symmetry, where $B$ and $L$ stand for Baryon and Lepton 
number, respectively. Now, one can have scenarios where $B-L$ is a local symmetry or 
scenarios where $B-L$ is not a local symmetry at the low-scale. At the same time there are 
several appealing ideas for physics beyond the Standard Model (SM) such as Supersymmetry 
and Grand Unification which deserve our attention. Then, it is important to understand the origin 
of neutrino masses in theories where SUSY is present or not, and in the context of 
grand unified theories, where one can understand the origin of the SM interactions 
and the correlation between fermion masses. It is easy to understand that combining 
all these ideas one finds different interesting theoretical frameworks and we should 
investigate the possible predictions that one could test at future neutrino experiments, 
at the LHC, or in the context of a grand unified theory, one should 
investigate the predictions for proton decay.     

The SM fermionic spectrum is very peculiar. We do not understand why the electron 
mass is much smaller than the top mass, and in general it is difficult to explain 
the hierarchies between the charged fermion masses. Now, the neutrino is the only 
neutral fermion in the SM and today thanks to the effort of many experimental 
collaborations we know that they are massive. We believe that the explanation 
of the fermion hierarchies demands the existence of physics beyond the SM, 
and in particular since the neutrino masses are so tiny, $m_\nu \sim 1 $ eV, 
perhaps they are special and the mechanism needed to explain their masses 
is different. 

In the neutrino sector one defines the so-called PMNS mixing matrix, 
$V_{PMNS}=V(\theta_{12},\theta_{13},\theta_{23}, \delta)$, where 
$\theta_{ij}$ and $\delta$ are the different mixing angles and the Dirac phase, 
respectively. In general there are two more free phases in the case of Majorana 
neutrinos. Thanks to all experimental collaborations one has very good constraints 
on the mixing angles and the mass squared difference, 
$\Delta m_{21}^2 = (7.2-8.9) \times 10^{-5}$ eV$^2$, $|\Delta m_{32}^2| = (2.1-3.1) 
\times 10^{-3}$ eV$^2$, $30^\circ < \theta_{12} < 38^\circ$, $ 36^\circ <  \theta_{23} < 54^\circ $, and $ \theta_{13} < 10^\circ$. 
Unfortunately, still we do not know if the spectrum for neutrinos has a 
Normal Hierarchy (NH) , Inverted Hierarchy (IH) or is Quasi-Degenerate (QD).   

Now, let us start with the properties of the neutrinos in the Standard Model.
As it is well-known the neutrinos are massless in the SM due to the conservation 
of the lepton number in each family, i.e. $U(1)_{L_{i}}$, with $L_i=L_e,L_\mu,L_\tau$, 
are accidental global symmetries. In general the neutrinos can be a 
\textit{Dirac} or \textit{Majorana} fermions. In the Dirac case one has to 
introduce a SM singlet, $\nu^C \sim (1,1,0)$, and the relevant interaction 
is given by
\begin{equation}
- {\cal L}^D_\nu = Y_\nu \ l \ H \ \nu^C \ + \ \textrm{h.c.}, 
\end{equation}
where $l^T=(\nu, e)_L$ and $H^T=(H^+, H^0)$ are the leptonic doublet and the 
Higgs, respectively. Then, in this case after electroweak symmetry breaking (EWSB) 
the neutrino mass matrix reads as: $M_\nu^D = Y_\nu \ v_0/\sqrt{2}$, 
with $v_0/\sqrt{2}$ the vacuum expectation value (vev) of the SM 
Higgs. Then, $Y_{\nu}$ should be around $10^{-11}$ in order 
to reproduce the correct neutrino mass ``scale", $m_\nu \sim 1$ eV.  This scenario 
is possible, however one has to \textit{impose by hand} the conservation 
of the total lepton number. Now, if higher-dimensional operators in the SM  
are allowed one expects that the neutrinos are naturally Majorana 
fermions since one finds the dimension five operator~\cite{Weinberg}:   
\begin{equation}
- {\cal L}^M_\nu = c_\nu \ (l \ H)^2 \ / \ \Lambda_\nu \ + \ \textrm{h.c.},
\label{Majorana}
\end{equation}
where $\Lambda_\nu$, typically called as the seesaw scale, corresponds 
to the scale where $L$ is broken. After EWSB one finds that 
$M_\nu^M= c_\nu \ v^2_0 / 2 \Lambda_\nu$. Now, if one assumes that the unknown 
coefficient $c_\nu$ is of order one the scale $\Lambda_\nu \approx 10^{14-15}$ GeV 
in order to reproduce the neutrino scale. Then, one could think that physics 
needed to explain neutrino masses is connected to the idea of grand 
unification since the unification scale is $M_{GUT} \sim 10^{14-16}$ GeV.
However, since in general the coefficient $c_\nu$ is a free parameter 
one could have the case where the scale $\Lambda_\nu$ is close to 
the electroweak scale. It is important to say that this possibility is appealing 
since one can hope to test directly this idea at the LHC or at future 
collider experiments. Now, what is the origin of the operator in Eq.(\ref{Majorana})?
There are many possible scenarios where one could understand the 
origin of this operator and those will be discussed in the next section. 
\section{Mechanisms for Neutrino Masses}
Let us discuss the simplest mechanisms for generating neutrino masses 
at tree level and one-loop level. The simplest mechanisms at tree level 
are the following:

{\bf \textit{\underline{Type I Seesaw}}}~\cite{TypeI}: 
This is perhaps the simplest mechanism for generating neutrino masses. 
In this case one adds a SM singlet, $\nu^C \sim (1,1,0)$, and using the interactions:    
\begin{equation}
- {\cal L}_\nu^I = Y_\nu \ l \ H \ \nu^C \ + \ \frac{1}{2} M \ \nu^C \ \nu^C \ + \ \textrm{h.c.}, 
\end{equation}
in the limit $M \gg Y_\nu v_0 $ one finds
\begin{equation}
{\cal M}_\nu^I =  \frac{1}{2} \ Y_\nu \ M^{-1} \ Y_\nu^T \ v^2_0,
\end{equation}
where M is typically defined by the $B-L$ breaking scale.
Then, one understands the smallness of the neutrino masses 
due to the existence of a mass scale, $M \gg Y_\nu v_0 \gg m_\nu$. 
Here, again if we assume $Y_\nu \sim 1$ the scale $M \sim 10^{14-15}$ GeV.
Now, in general it is not possible to make predictions for the neutrino 
masses and mixing in this framework since we do not know the 
matrices $Y_\nu$ and $M$. Then, one should look for a theory 
where one could predict these quantities. 

{\bf \textit{\underline{Type II Seesaw}}}~\cite{TypeII}: 
In this scenario one introduces a new Higgs boson, 
$\Delta \sim (1,3,1)$, which couples to the leptonic 
doublets and the SM Higgs boson:
\begin{equation}
{\cal L}_\nu^{II}= - Y_\nu \ l \ \Delta \ l \ + \ \mu \ H \  \Delta^\dagger \ H \ + \ \textrm{h.c.},
\end{equation}
and when the neutral component in $\Delta=(\delta^0, \delta^+, \delta^{++})$ gets a vev, $v_\Delta$, 
one finds:
\begin{equation}
\label{TypeII}
{\cal M}_\nu^{II}= \sqrt{2} \ Y_\mu \ v_{\Delta} = \mu \ Y_\nu \ v_0^2 / M_{\Delta}^2. 
\end{equation}
Notice that if $\mu \sim M_{\Delta}$ and $M_{\Delta} \sim 10^{14-15}$ GeV 
the vev $v_{\Delta}$ should be of order $1$ eV. However, 
in general the triplet mass can be around the TeV scale 
and $\mu$ can be small. 
Now, one should know the matrix $Y_\nu$, $\mu$ 
and $M_{\Delta}$ in order to make predictions 
for neutrino mixing and masses. Then, as in 
the previous case, one should look for a theory 
where one can predict these quantities.  

Here I cannot discuss the testability of seesaw mechanisms 
at the LHC since this topic is beyond the scope of this talk 
but I would like to mention an interesting scenario. Suppose 
that $M_{\Delta} \leq 1$ TeV and $v_{\Delta} < 10^{-4}$ GeV.
In this case one could produce at the LHC the doubly and singly 
charged Higgses present in the model and through the dominant decays,
$H^{++} \to e_i^+ e^+_j$ and $H^{+} \to e_i^+ \bar{\nu}$, 
we could learn about the neutrino spectrum. In Fig. 1 one can see 
the predictions for the branching ratios of $H^{++}$ versus the 
lightest neutrino mass~\cite{TypeII-LHC}.
\begin{figure}[tb]
\includegraphics[scale=1,width=7.0cm]{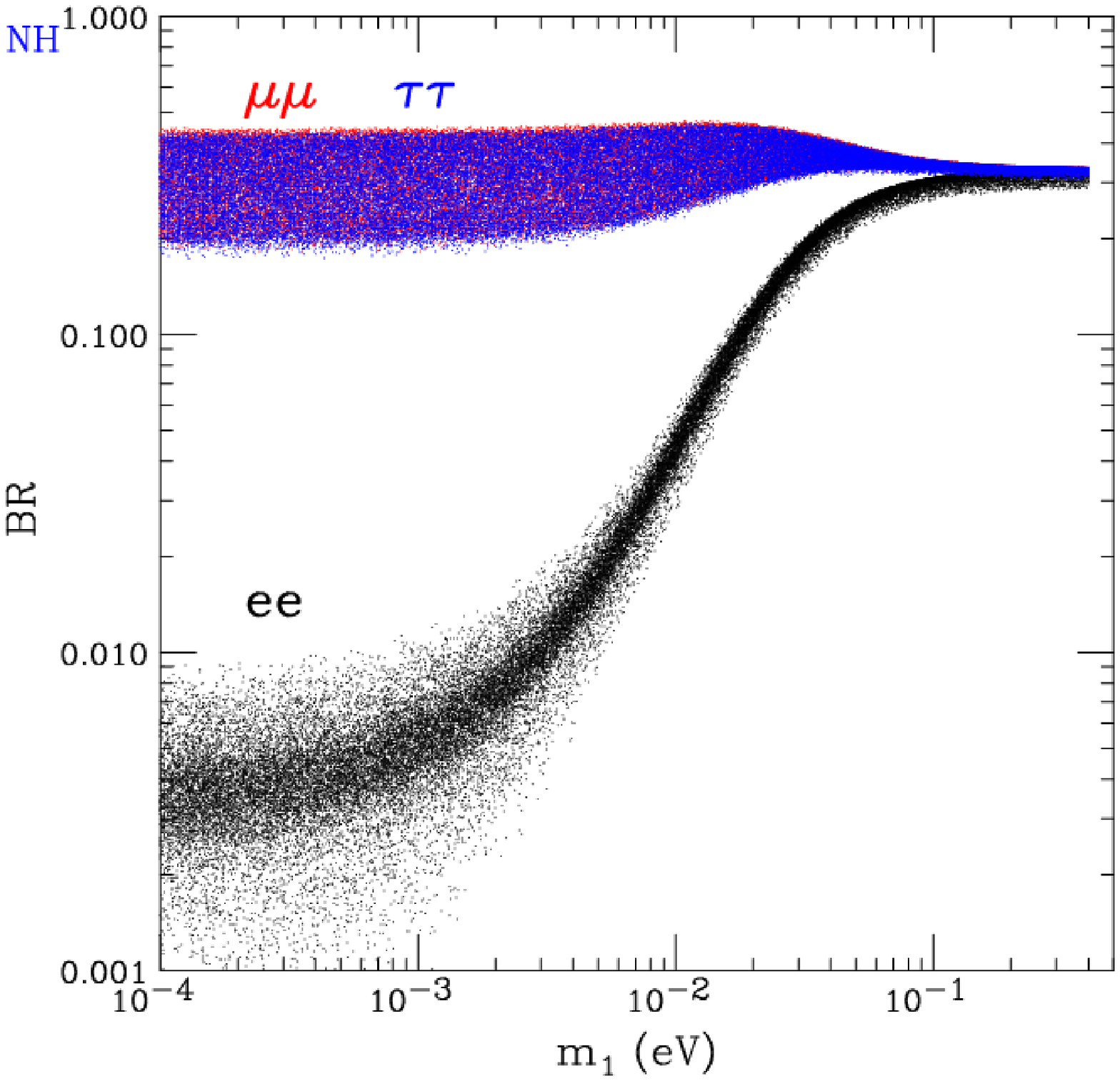}
\includegraphics[scale=1,width=7.0cm]{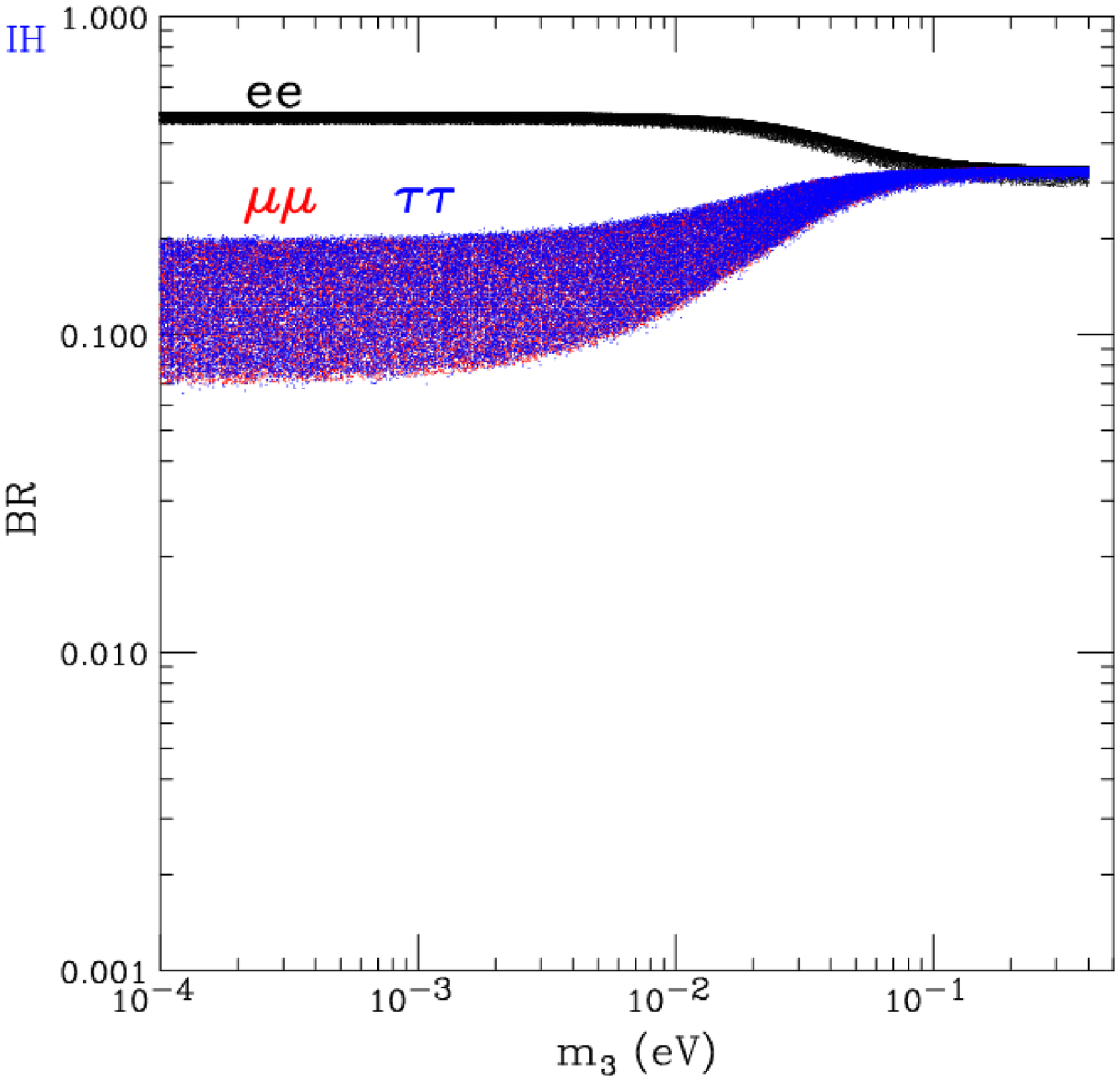}
\caption{Scatter plots for the $H^{++}$ decay branching fractions 
versus the lowest neutrino mass for NH (left) and IH (right).}
\label{brii}
\end{figure}
Notice that using the properties of the doubly charged Higgs 
decays in each spectrum one can distinguish between NH, 
IH or QD. Now, in the case when the Majorana phases play 
an important role the predictions in Fig. 1 can change dramatically, 
and in order to learn about the neutrino spectrum it is better 
to use the singly charged Higgs decays~\cite{TypeII-LHC}. 
Then, as it has been proposed in Ref.~\cite{TypeII-LHC}, the associated 
production $H^{\pm \pm} H^{\mp}$, is crucial for the test 
of the mechanism and to learn about the spectrum for neutrinos. 
For other studies see Ref.~\cite{TypeII-LHC-Others}.

{\bf \textit{\underline{Type III Seesaw}}}~\cite{TypeIII,Ma,Goran,Adjoint-SU(5),SUSY-Adjoint-SU(5),LR-TypeIII}:
In the case of Type III seesaw one adds new 
fermions, $\rho \sim (1,3,0)$, and the neutrino 
masses are generated using the following interactions:
\begin{equation}
- {\cal L}_\nu^{III}= Y_\nu \ l \ \rho \ H \ + \ M_{\rho} \ \textrm{Tr} \  \rho^2 \ + \ \textrm{h.c.},
\end{equation}
where $\rho=(\rho^0, \rho^+, \rho^-)$. Integrating 
out the neutral component of the fermionic triplet 
one finds 
\begin{equation}
{\cal M}_\nu^{III}= \frac{1}{2} Y_\nu \ M_{\rho}^{-1} \ Y_\nu^T \ v_0^2.
\end{equation}
Here, as in the case of Type I seesaw, if $Y_\nu \sim 1$ one needs 
$M_{\rho} \sim 10^{14-15}$ GeV. Here one 
faces the same problem, if we want to make predictions for 
neutrinos masses and mixings, a theory where $Y_\nu$ 
and $M_{\rho}$ can be predicted is needed. In the next 
section we will discuss this issue in the context of 
grand unified theories.

We have mentioned the simplest mechanisms at tree level.
Now, if Supersymmetry is realized in nature one has the 
extra possibility to generate neutrino masses through 
the R-parity violating couplings. R-parity is defined 
as $R=(-1)^{3(B-L) + 2 S}$, and the R-parity violating 
interactions are given by
\begin{eqnarray}
{\cal W}_{RpV}&=& \epsilon_i \ \hat{L}_i \ \hat{H}_u 
\ + \  \lambda_{ijk} \ \hat{L}_i \ \hat{L}_j  \ \hat{E}_k^C \ + \ \lambda_{ijk}^{'} \ \hat{Q}_i \ \hat{L}_j \ \hat{D}^C_k 
\nonumber \\
& + & \lambda_{ijk}^{''} \ \hat{U}^C_i \ \hat{D}^C_j \ \hat{D}^C_k,
\end{eqnarray}
where the last term violates B and the others break L. The problem 
with this possibility is that in general one has too many 
free parameters. Then, it is also important to understand 
the origin of these interactions in a theory where $R$-parity 
is spontaneously broken. Recently, this issue has been 
investigated in Ref.~\cite{Sogee} in the context of different 
theories where $B-L$ is a local symmetry.

There are several mechanisms for generating neutrino 
masses at one-loop level. In this case one assumes 
that the mechanisms discussed above are absent and 
only though quantum corrections one generates 
neutrino masses. This possibility is very appealing 
since the neutrino masses are very tiny and the 
seesaw ``scale" can be low.

{\bf \textit{\underline{Zee Model}}}~\cite{Zee}: 
In the so-called Zee model one introduces two extra 
Higgs bosons, $h\sim (1,1,1)$ and $H^{'} \sim (1,2,1/2)$.
In this case the relevant interactions are
\begin{equation}
- {\cal L}_{Zee}= Y \ l \ h\ l \ + \ \mu \ H \ H^{'} \ h^\dagger 
\ + \  \sum_{i=1}^2 \ Y_i \ e^C \ H^\dagger_i  \ l \ + \ \textrm{h.c.},  
\end{equation}
where in general both Higgs doublets couple to the matter fields.
Using these interactions one can generate neutrino masses 
at one-loop level. See Ref.~\cite{Zee} for details. Now, 
it is important to mention that in the simple case where only 
one Higgs doublet couples to the leptons~\cite{Wolfenstein} 
it is not possible to generate neutrino masses in 
agreement with neutrino data. See for example 
Ref.~\cite{He} for details.

{\bf \textit{\underline{A New Mechanism at One-Loop Level}}}~\cite{Fileviez-Perez-Wise}:
Now, suppose that one looks for the simplest mechanism for neutrino 
masses at one-loop level where we add only two types of representations, 
a fermionic and a scalar one, and with no extra symmetry. All the possibilities 
were considered in Ref.~\cite{Fileviez-Perez-Wise} where we found that only two cases 
are allowed by cosmology. In this case one has two possible 
cases: 1) The extra fields are a fermionic $\rho_1 \sim (8,1,0)$ 
and the scalar $S \sim (8,2,1/2)$. 2) One adds 
$\rho_2 \sim (8,3,0)$ and $S \sim (8,2,1/2)$. In both cases one 
generates neutrinos masses through the loop in Fig. 2.

\begin{center}
\begin{picture}(200,120)(0,0)
\ArrowLine(20,0)(60,0)
\ArrowLine(100,0)(60,0)
\ArrowLine(100,0)(140,0)
\ArrowLine(180,0)(140,0)
\DashArrowLine(68,72)(100,40)6
\DashArrowLine(132,72)(100,40)6
\DashArrowArc(100,0)(40,90,180)6
\DashArrowArcn(100,0)(40,90,0)6
\Text(40,-8)[c]{$\nu_i$}
\Text(80,-8)[c]{$\rho$}
\Text(100,0)[c]{$\times$}
\Text(120,-8)[c]{$\rho$}
\Text(160,-8)[c]{$\nu_j$}
\Text(59,28)[c]{$S$}
\Text(142,28)[c]{$S$}
\Text(60,80)[c]{$H^0$}
\Text(140,80)[c]{$H^0$}
\end{picture}
\vskip 0.2in
{\bf Fig.~2.} ~ New mechanism at one-loop level.
\end{center}

The relevant interactions in this case are given by
\begin{equation}
-{\cal L} = Y_2 \ l \ S \  \rho_1 \ + \ M_{\rho_1} \ {\rm Tr} \ \rho^2_1  
\ + \ \lambda_2 \  {\rm Tr} \left( S^\dagger H \right)^2 \ + \ {\rm h.c.} ~.
\label{V2}
\end{equation}
Using as input parameters, $M_{\rho_1} = 200$ GeV, $v_0=246$ GeV 
and $M_{S}=2$ TeV we find that in order to get 
the neutrino ``scale", $\sim 1$ eV, the combination 
of the couplings, $Y_2^2 \lambda_2 \sim 10^{-8}$. 
This mechanism could be tested at the LHC through 
the channels $pp \ \to \ \rho_1 \rho_1 \ \to \ S^+ S^+  e^-_i   e^-_j \ \to \ e^-_i e^-_j t t \bar{b} \bar{b}$
~\cite{Fileviez-Perez-Wise}. See Ref.~\cite{Sean} for the study of leptogenesis in this context. 
\section{Grand Unification and Massive Neutrinos}
The so-called grand unified theories are one of the most appealing 
extensions of the SM where one can understand the origin of SM 
interactions. Here we will discuss the implementation of the different 
mechanisms for neutrino masses in the context of renormalizable 
$SU(5)$ and $SO(10)$ theories.

{\bf \textit{\underline{$SU(5)$ and Neutrino Masses}}}:
The original model proposed by Georgi and Glashow~\cite{GG} 
in 1974 has been considered as the simple grand unified 
theory. This model is based on $SU(5)$, the SM matter fields 
live in the $\overline{5}=(d^C, l)$ and $10=(u^C, Q, e^C)$ 
representations, and the minimal Higgs sector is composed 
of $5_H$ and $24_H$. As is well-known this model 
is ruled out by unification. At the same time one has 
$M_D=M_E^T$ which is in disagreement with the experiment 
and there are no neutrino masses. In order to have a consistent 
relation between the masses of down quarks and charged 
leptons one has two possibilities: a) one introduces 
a  $45_H$~\cite{45}, b) one includes higher-dimensional operators~\cite{Ellis}.
In the case of  neutrinos masses one can have the 
mechanisms at tree level mentioned above:  
i) we can introduces at least two singlets and use the 
Type I seesaw, ii) in the case of Type II seesaw one needs 
to introduces a new Higgs $15_H$ ($\hat{15}_H$ and 
$\hat{\overline{15}}_H$ in the SUSY case), iii) a new 
fermionic $24$ representation is needed to realize 
the Type III seesaw mechanism. Since the simplest $SU(5)$ 
model with Type I seesaw is ruled out by unification I would 
like to focus on the models with Type II or Type III seesaw.

\begin{figure}[tb]
\includegraphics[scale=1,width=9.0cm]{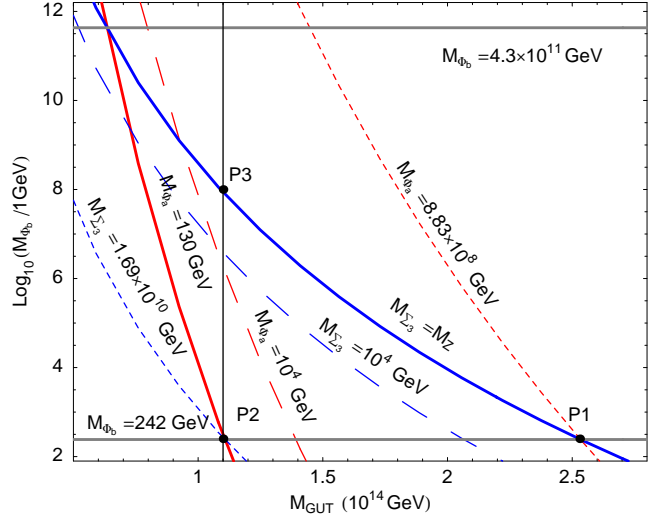}
\caption{Full parameter space allowed by unification at one-loop 
level in the model proposed in~\cite{TypeII-SU(5)}. For details see Ref.~\cite{German}.}
\label{triangle}
\end{figure}

{\bf \textit{Type II-SU(5)}}~\cite{TypeII-SU(5)}: One can realize a simple realistic 
$SU(5)$ theory when the neutrino masses are generated 
through the Type II seesaw mechanism. In this case the 
Higgs sector is composed of $5_H$, $15_H$ and $24_H$
~\cite{TypeII-SU(5)} and one can have unification in  
agreement with proton decay lifetime bounds 
and all experimental constraints. See Refs.~\cite{Ricardo} 
and~\cite{German} for details. Now, the 
$15_H=(\Phi_a,\Phi_b,\Phi_c)=(1,3,1)\bigoplus(3,2,1/6)\bigoplus(6,1,-2/3)$ 
contains the field needed for seesaw $\Phi_a=i \sigma_2 \Delta$ and 
relevant interactions are
\begin{equation}
V_\nu = Y_\nu \ \overline{5} \  \overline{5} \ 15_H \ + \  \mu \ 5_H^* \ 5_H^* \ 15_H \ + \ \textrm{h.c.}. 
\end{equation}   
In this case the mass matrix for neutrinos is given by Eq.~(\ref{TypeII}). 
It is clear from Eq.~(\ref{TypeII}) one cannot predict the neutrino 
masses and mixing. However, let us discuss the 
possible constraints on the seesaw scale in this case. 
In Fig.~\ref{triangle} we show the full parameter space allowed 
by unification. Now, one can make two 
observations: a) the mass of the seesaw triplet,  
$\Phi_a=i \sigma_2 \Delta$, has to be in the range 
$100 \ \textrm{GeV}  \  \leq \  M_{\Delta} \  \leq \ 9 \times 10^8$ GeV. 
Then, one can say that the seesaw scale in this context 
can be very low in a consistent way. Maybe, this is a good 
way to justify the studies in Ref.~\cite{TypeII-LHC}. 
b) If one studies results shown in Fig.~\ref{triangle} 
it is easy to see that the leptoquark, $\Phi_b \sim (3,2,1/6)$,  
is very light in a large region of the parameter space. 
Then, this is perhaps a way to test this theory at colliders.
See Ref.~\cite{Leptoquark-LHC} for the study of this 
leptoquark signatures at the LHC. 
 
\begin{figure}[tb]
\includegraphics[scale=1,width=9.0cm]{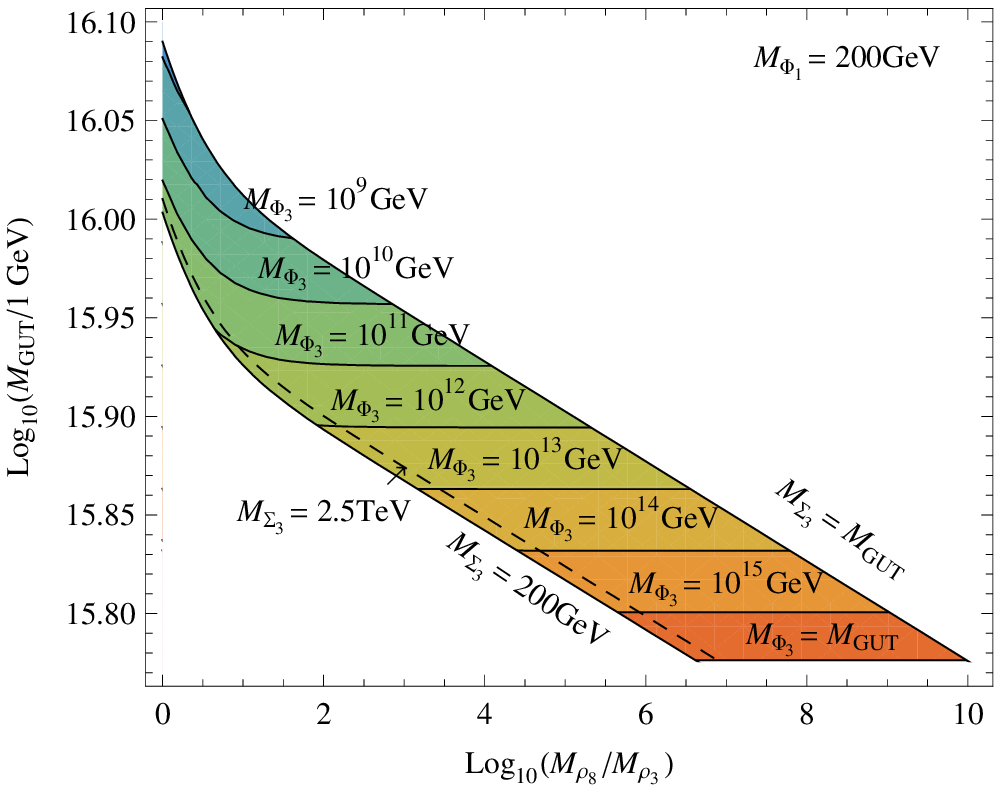}
\caption{Full parameter space allowed by unification at one-loop 
level in Adjoint SU(5)~\cite{Adjoint-SU(5)} when the color octet is light. 
For details see Ref.~\cite{German-Hoernisa}.}
\label{Adjoint}
\end{figure}

{\bf \textit{Adjoint $SU(5)$}}: The implementation of the Type III seesaw 
mechanism in the context of grand unified theories have been 
studied by several groups~\cite{Ma,Goran,Adjoint-SU(5),SUSY-Adjoint-SU(5)}.
Here we will focus on the first realization of the mechanism 
in a renormalizable GUT model~\cite{Adjoint-SU(5),SUSY-Adjoint-SU(5)}, 
we refer to this theory as Adjoint-SU(5). In this theory the matter 
fields live in $\overline{5}$, $10$ and $24$, while the Higgs sector 
is composed of $5_H$, $24_H$ and $45_H$.  Once one has the 
decomposition of  $24=(\rho_8,\rho_3,\rho_{(3,2)},\rho_{(\bar{3},2)},\rho_0)=(8,1)\bigoplus(1,3)\bigoplus(3,2)\bigoplus(\bar{3},2)\bigoplus(1,1)$ it is easy to realize that the mechanism 
for neutrino masses is a combination of Type I and Type III seesaw.
The relevant interactions for our discussion are:
\begin{eqnarray}
V_\nu &=& c_\alpha \ \bar{5}_\alpha \ 24 \ 5_H \ + \ p_\alpha \ \bar{5}_\alpha \ 24 \ 45_H \nonumber \\
& + & M \ \textrm{Tr} \ 24^2 \ + \ \lambda \ \textrm{Tr} \left( 24^2 24_H \right) \ + \ \textrm{h.c.}.
\label{Adjoint-Potential}
\end{eqnarray}
Now, integrating out the singlet, $\rho_0$, and the neutral component of the triplet, 
$\rho_3$, one finds that the mass matrix for neutrinos is given by
\begin{equation}
{\cal M}^\nu_{\alpha \beta} = \frac{h_{\alpha 1} \ h_{\beta 1}}{ M_{\rho_0}} \ v_0^2 \ + \ 
\frac{h_{\alpha 2} \ h_{\beta 2}}{ M_{\rho_3}} \ v_0^2.
\end{equation} 
Then, we have as prediction one massless neutrino and the 
spectrum can be: $m_1=0$, $m_2=\sqrt{\Delta m_{sol}^2}$ 
and $m_3=\sqrt{\Delta m_{sol}^2 \ + \ \Delta m_{atm}^2}$ in 
the case of  NH  or $m_3=0$, $m_2=\sqrt{\Delta m_{atm}^2}$, 
and $m_1=\sqrt{\Delta m_{atm}^2 \ - \  \Delta m_{sol}^2 }$ 
in the case of IH. Here $\Delta m^2_{sol} \approx 8 \times 10^{-5}$ eV$^2$ and 
 $\Delta m^2_{atm} \approx 2.5 \times 10^{-3}$ eV$^2$ are the 
 solar and atmospheric mass squared differences. In order to prove 
 that in this model one can satisfy all constraints coming from proton 
 decay searches we show in Fig.~\ref{Adjoint} the possible 
 predictions coming from the unification of gauge couplings.
It is important to make several observations in this case: 
a) In this model the seesaw triplet can be light only 
if we have a fine-tuning between the last two terms 
in Eq.~(\ref{Adjoint-Potential}), b) In order to 
have gauge unification in agreement with proton decay 
bounds one should have a light color octet. 
See Ref.~\cite{Color-Octet-LHC} for the study 
of color octets at the LHC. c) This model could be 
tested at future proton decay experiments since 
the upper bounds on the lifetimes are:  $\tau (p \to K^+ \bar{\nu}) \leq 10^{37}$ 
years and $\tau (p \to \pi^+ \bar{\nu}) \leq  3 \times 10^{35}$ years.  
For the study of the leptogenesis mechanism in this context 
see Ref.~\cite{Steve}.
In our opinion these are the simplest models based on $SU(5)$ 
where one could to understand the origin of neutrino masses.
Of course, one can add in those models a flavour symmetry 
and study the predictions for neutrino mixings but this issue is 
beyond the scope of this letter.

\begin{figure}[h]
\includegraphics[scale=1,width=15.0cm]{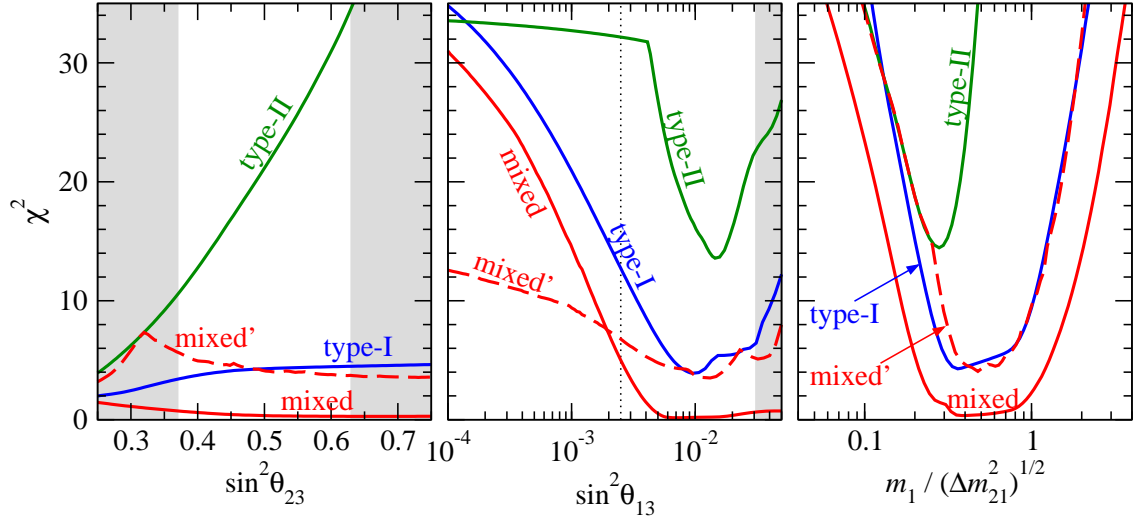}
\caption{Predictions for the neutrino mixings in $SO(10)$ models 
where the Higgs sector is composed of $10_H$, $126_H$, and $\overline{126}_H$.
For details see Ref.~\cite{Bertolini}.}
\label{Bertolini}
\end{figure}
{\bf \textit{\underline{Renormalizable SO(10) and Neutrino Masses}}}:
Now, let us study the neutrino mass mechanisms in grand unified theories 
based on $SO(10)$~\cite{SO(10)}. For a review on $SO(10)$ see 
Ref.~\cite{Goran-SO(10)}. Here, I will focus on the study of renormalizable 
theories since we would like to know the possible predictions for 
neutrinos in the case when we stick only to the idea of grand unified 
theories. In SO(10) one has the possibility to unify all matter fields 
of  one family in the spinor representation $16=(Q,u^C,d^C,L,e^C,\nu^C)$.
Now, since one can have the right-handed neutrino in $16$ one expects 
from the beginning that the neutrino masses will be generated at least 
through the Type I seesaw mechanism. In naive $SO(10)$ one can generate 
fermion masses using the interactions
\begin{equation}
-{\cal L}_Y = Y_{10} \ 16 \ 16 \ 10_H  \ + \  \textrm{h.c.},
\end{equation}
and one finds the following relations
\begin{eqnarray}
M_U &=& M^D_\nu = v^u_{10} \ Y_{10}  \  \  (\textrm{wrong}) , \\
M_D &=& M_E = v^d_{10}  \ Y_{10} \  \  (\textrm{wrong}),  \\ 
Y_{10} &=& Y^T_{10}.
\end{eqnarray}
As one can see the minimal $SO(10)$ model fails 
badly since one cannot have a consistent relation 
for fermion masses. Unfortunately, in order to 
realize a realistic SUSY model at the renormalizable 
level one needs to introduce a new large Higgs 
representation, $126_H$ and $\overline{126}_H$.
In this case the relevant Yukawa interactions are
\begin{equation}
-{\cal L}_Y^{R-SO(10)} = Y_{10} \ 16 \ 16 \ 10_H  
\ + \  Y_{126} \ 16 \ 16 \ \overline{126}_H \ + \ \textrm{h.c.},
\end{equation}
and one finds the relations
\begin{eqnarray}
{\cal M}_U &=& v_{10}^u \ Y_{10} \ + \ v_{126}^u \ Y_{126}, \\
{\cal M}^D_\nu  &=& v^u_{10} \  Y_{10} \ - \ 3  \ v^u_{126} \  Y_{126}, \\
{\cal M}_D &=& v^d_{10} \  Y_{10} \ + \ v^d_{126} \  Y_{126}, \\
{\cal M}_E &=& v^d_{10} \  Y_{10} \ - \ 3 \ v^d_{126} \  Y_{126}, \\
{\cal M}_{\nu_R} &=& Y_{126} \  v_R, \\
{\cal M}_\nu &=& - {\cal M}^D_\nu \  {\cal M}^{-1}_{\nu_R} \  {\cal M}^D_\nu  \ + \ Y_{126} \ v_L.
\end{eqnarray} 
Now, as one can appreciate in this context the 
neutrino masses are generated through the Type I and Type II 
seesaw mechanisms and taking as input parameters all 
experimental values for charged fermion masses and mixings 
one can make predictions in the neutrino sector.  
See Refs.~\cite{Fermion-masses-SO(10),Bertolini} 
for the study of fermion masses in this context. 
In order to illustrate the possible predictions in this 
context we will take as an example the results shown 
in Ref.~\cite{Bertolini}. In Fig.~\ref{Bertolini} 
we show possible predictions for the neutrino 
mixing angles. In this case only when one assumes 
a particular scenario for the seesaw mechanism 
we can talk about a prediction for the mixing angles.
For example, in the case of $\theta_{13}$ one 
can have a better fit in the mixed scenario, where 
one has the type I and Type II contributions, 
and the prefered value for $\sin^2 \theta_{13}$ 
is around $10^{-2}$.  There are many aspects 
of these models that we cannot cover here 
and we refer the reader to Ref.~\cite{Fermion-masses-SO(10),Bertolini}.
For the possible predictions in models with additional 
flavour symmetries see Ref.~\cite{Lindner}.

\section{Summary}
We have discussed the simplest models for the generation of neutrinos masses at tree level 
and one-loop level. The realization of the different seesaw mechanisms in the context 
of renormalizable $SU(5)$ and $SO(10)$ theories and possible predictions 
for neutrino masses and mixing have been briefly reviewed. A new mechanism 
for the generation of neutrino masses at one-loop level was presented. 
We discussed the first realization of the Type III seesaw mechanism 
in the context of a renormalizable $SU(5)$ theory, called ``Adjoint SU(5)".  

\begin{theacknowledgments}
I would like to thank the organizers of NuFact09 for the invitation 
and for this enjoyable and very well organized meeting in Chicago.
I would like to thank my collaborators V. Barger, S. Blanchet, 
M. A. Diaz, I. Dorsner,  L. Everett, R. Gonzalez Felipe, 
T. Han, G.Y. Huan, H. Iminniyaz, T. Li, G. Rodrigo, 
G. Senjanovi\'c, S. Spinner, K. Wang, and M. B. Wise
for many discussions and enjoyable collaboration. 
This work was supported in part by the U.S. Department 
of Energy contract No. DE-FG02-08ER41531 
and in part by the Wisconsin Alumni Research Foundation.  
Last, but not least, I would like to thank the White Sox 
baseball team for winning a perfect game while I was 
visiting Chicago.
\end{theacknowledgments}
\newpage


\begin{thebibliography}{9}
\bibitem{Weinberg}
  S.~Weinberg,
  Phys.\ Rev.\ Lett.\  {\bf 43}, 1566 (1979).

\bibitem{TypeI}
 P.~Minkowski,
  Phys.\ Lett.\  B {\bf 67}, 421 (1977);
  T. Yanagida, 
   (KEK Report~79-18, Tsukuba, 1979), p.~95;
  M. Gell-Mann, P. Ramond and R. Slansky,
   in {\it Supergravity}, eds. P. van Nieuwenhuizen et al.,
   (North-Holland, 1979), p.~315;
  S.L. Glashow, in {\it Quarks and Leptons}, Carg\`ese, eds. M. L\'evy et al.,
(Plenum, 1980), p. 707;
  R.~N.~Mohapatra and G.~Senjanovic,
  Phys.\ Rev.\ Lett.\  {\bf 44}, 912 (1980).
 
 \bibitem{TypeII}
 W.~Konetschny and W.~Kummer, 
  Phys.\ Lett.\  B {\bf 70} (1977) 433;
T.~P.~Cheng and L.~F.~Li,
  Phys.\ Rev.\  D {\bf 22} (1980) 2860;
  G.~Lazarides, Q.~Shafi and C.~Wetterich,
  Nucl.\ Phys.\ B {\bf 181} (1981) 287;
  J.~Schechter and J.~W.~F.~Valle,
  Phys.\ Rev.\ D {\bf 22} (1980) 2227;
  R.~N.~Mohapatra and G.~Senjanovi\'c,
  Phys.\ Rev.\ D {\bf 23} (1981) 165.
  
\bibitem{TypeII-LHC}
  P.~Fileviez P\'erez, T.~Han, G.~Y.~Huang, T.~Li and K.~Wang,
  Phys.\ Rev.\  D {\bf 78} (2008) 071301;
  Phys.\ Rev.\  D {\bf 78}, 015018 (2008).


\bibitem{TypeII-LHC-Others}
See Talk by F. del Aguila at NuFact09.
See also:
  J.~Garayoa and T.~Schwetz,
  JHEP {\bf 0803}, 009 (2008);
  A.~G.~Akeroyd, M.~Aoki and H.~Sugiyama,
  Phys.\ Rev.\  D {\bf 77}, 075010 (2008);
  F.~del Aguila and J.~A.~Aguilar-Saavedra,
  Nucl.\ Phys.\  B {\bf 813}, 22 (2009).
  
\bibitem{TypeIII}   
R.~Foot, H.~Lew, X.~G.~He and G.~C.~Joshi,
  Z.\ Phys.\ C {\bf 44} (1989) 441. 
  
\bibitem{Ma}
E.~Ma,
  Phys.\ Rev.\ Lett.\  {\bf 81} (1998) 1171;

\bibitem{Goran}
B.~Bajc and G.~Senjanovi\'c,
  JHEP {\bf 0708} (2007) 014;
  
 \bibitem{Adjoint-SU(5)} 
  P.~Fileviez~P\'erez,
  Phys.\ Lett.\  B {\bf 654} (2007) 189.
  
\bibitem{SUSY-Adjoint-SU(5)}  
  P.~Fileviez~P\'erez,
  Phys.\ Rev.\  D {\bf 76} (2007) 071701.

\bibitem{LR-TypeIII}
 P.~Fileviez P\'erez,
  JHEP {\bf 0903} (2009) 142.
 
\bibitem{Sogee}
  P.~Fileviez P\'erez and S.~Spinner,
  Phys.\ Lett.\  B {\bf 673} (2009) 251;
  Phys.\ Rev.\  D {\bf 80} (2009) 015004;
  V.~Barger, P.~Fileviez P\'erez and S.~Spinner,
  Phys.\ Rev.\ Lett.\  {\bf 102} (2009) 181802;
  L.~L.~Everett, P.~Fileviez~P\'erez and S.~Spinner,
  Phys.\ Rev.\  D {\bf 80} (2009) 055007.

\bibitem{Zee}
  A.~Zee,
  Phys.\ Lett.\  B {\bf 93} (1980) 389
  [Erratum-ibid.\  B {\bf 95} (1980) 461].
 
\bibitem{Wolfenstein}
  L.~Wolfenstein,
  Nucl.\ Phys.\  B {\bf 175} (1980) 93.

  \bibitem{He}
  P.~H.~Frampton and S.~L.~Glashow,
  Phys.\ Lett.\  B {\bf 461} (1999) 95;
  P.~H.~Frampton, M.~C.~Oh and T.~Yoshikawa,
  Phys.\ Rev.\  D {\bf 65} (2002) 073014;
  X.~G.~He,
  Eur.\ Phys.\ J.\  C {\bf 34} (2004) 371;
  A.~Y.~Smirnov and M.~Tanimoto,
  Phys.\ Rev.\  D {\bf 55} (1997) 1665;
  Y.~Koide,
  Phys.\ Rev.\  D {\bf 64} (2001) 077301.

\bibitem{Fileviez-Perez-Wise}
    P.~Fileviez P\'erez and M.~B.~Wise,
  Phys.\ Rev.\  D {\bf 80} (2009) 053006.

\bibitem{Sean}
  M.~Losada and S.~Tulin,
  arXiv:0909.0648 [hep-ph].

\bibitem{GG}
  H.~Georgi and S.~L.~Glashow,
  Phys.\ Rev.\ Lett.\  {\bf 32} (1974) 438.
 
\bibitem{45}
  H.~Georgi and C.~Jarlskog,
  Phys.\ Lett.\  B {\bf 86} (1979) 297.
  
\bibitem{Ellis}
  J.~R.~Ellis and M.~K.~Gaillard,
  Phys.\ Lett.\  B {\bf 88} (1979) 315.
  
\bibitem{TypeII-SU(5)}
  I.~Dorsner and P.~Fileviez P\'erez,
  Nucl.\ Phys.\  B {\bf 723} (2005) 53.
  
\bibitem{Ricardo}
  I.~Dorsner, P.~Fileviez P\'erez and R.~Gonzalez Felipe,
  Nucl.\ Phys.\  B {\bf 747} (2006) 312.

\bibitem{German}
  I.~Dorsner, P.~Fileviez P\'erez and G.~Rodrigo,
  Phys.\ Rev.\  D {\bf 75} (2007) 125007.
       
\bibitem{Leptoquark-LHC}
  P.~Fileviez P\'erez, T.~Han, T.~Li and M.~J.~Ramsey-Musolf,
  Nucl.\ Phys.\  B {\bf 819} (2009) 139.
  
\bibitem{German-Hoernisa}
  P.~Fileviez P\'erez, H.~Iminniyaz and G.~Rodrigo,
  Phys.\ Rev.\  D {\bf 78} (2008) 015013.
 
\bibitem{Color-Octet-LHC} 
M.~I.~Gresham and M.~B.~Wise,
  Phys.\ Rev.\  D {\bf 76} (2007) 075003;
  P.~Fileviez P\'erez, R.~Gavin, T.~McElmurry and F.~Petriello,
  Phys.\ Rev.\  D {\bf 78} (2008) 115017;
    M.~Gerbush, T.~J.~Khoo, D.~J.~Phalen, A.~Pierce and D.~Tucker-Smith,
  Phys.\ Rev.\  D {\bf 77} (2008) 095003;
    A.~R.~Zerwekh, C.~O.~Dib and R.~Rosenfeld,
  Phys.\ Rev.\  D {\bf 77} (2008) 097703;
  A.~Idilbi, C.~Kim and T.~Mehen,
  [arXiv:0903.3668[ [hep-ph]];
  C.~P.~Burgess, M.~Trott and S.~Zuberi,
  JHEP {\bf 0909} (2009) 082.
  
\bibitem{Steve}
  S.~Blanchet and P.~Fileviez P\'erez,
  JCAP {\bf 0808} (2008) 037;
  Mod.\ Phys.\ Lett.\  A {\bf 24} (2009) 1399.
See also: 
  W.~Fischler and R.~Flauger,
  JHEP {\bf 0809} (2008) 020.
  
\bibitem{SO(10)}
H. Georgi, \textit{In Coral Gables 1979 Proceeding, Theory and experiments 
in high energy physics}, New York 1975;
  H.~Fritzsch and P.~Minkowski,
  Annals Phys.\  {\bf 93} (1975) 193.

\bibitem{Goran-SO(10)}
  G.~Senjanovi\'c,
  arXiv:hep-ph/0612312.
  
\bibitem{Fermion-masses-SO(10)}
  T.~E.~Clark, T.~K.~Kuo and N.~Nakagawa,
  Phys.\ Lett.\  B {\bf 115} (1982) 26;
  C.~S.~Aulakh and R.~N.~Mohapatra,
  Phys.\ Rev.\  D {\bf 28} (1983) 217;
  C.~S.~Aulakh, B.~Bajc, A.~Melfo, G.~Senjanovic and F.~Vissani,
  Phys.\ Lett.\  B {\bf 588} (2004) 196;
  K.~S.~Babu and R.~N.~Mohapatra,
  Phys.\ Rev.\ Lett.\  {\bf 70} (1993) 2845;
  B.~Brahmachari and R.~N.~Mohapatra,
  Phys.\ Rev.\  D {\bf 58} (1998) 015001;
  K.~Matsuda, Y.~Koide and T.~Fukuyama,
  Phys.\ Rev.\  D {\bf 64} (2001) 053015;
  K.~Matsuda, Y.~Koide, T.~Fukuyama and H.~Nishiura,
  Phys.\ Rev.\  D {\bf 65} (2002) 033008
  [Erratum-ibid.\  D {\bf 65} (2002) 079904];
  T.~Fukuyama and N.~Okada,
  JHEP {\bf 0211} (2002) 011;
  H.~S.~Goh, R.~N.~Mohapatra and S.~P.~Ng,
  Phys.\ Lett.\  B {\bf 570} (2003) 215;
  H.~S.~Goh, R.~N.~Mohapatra and S.~P.~Ng,
  Phys.\ Rev.\  D {\bf 68} (2003) 115008;
  S.~Bertolini and M.~Malinsky,
  Phys.\ Rev.\  D {\bf 72} (2005) 055021;
  K.~S.~Babu and C.~Macesanu,
  Phys.\ Rev.\  D {\bf 72} (2005) 115003;
  C.~S.~Aulakh,
  arXiv:hep-ph/0506291;
  B.~Bajc, A.~Melfo, G.~Senjanovic and F.~Vissani,
  Phys.\ Lett.\  B {\bf 634} (2006) 272;
  C.~S.~Aulakh and S.~K.~Garg,
  Nucl.\ Phys.\  B {\bf 757} (2006) 47;
  B.~Bajc, I.~Dorsner and M.~Nemevsek,
  JHEP {\bf 0811} (2008) 007.

\bibitem{Bertolini}
  S.~Bertolini, T.~Schwetz and M.~Malinsky,
  Phys.\ Rev.\  D {\bf 73} (2006) 115012.
  
\bibitem{Lindner}
See talks by M. Lindner 
and C. Albright at NuFact09.
For reviews see for example:
  G.~Altarelli,
  arXiv:0905.3265 [hep-ph];
  E.~Ma,
  arXiv:0908.1770 [hep-ph];
  M.~C.~Chen and K.~T.~Mahanthappa,
  Nucl.\ Phys.\ Proc.\ Suppl.\  {\bf 188} (2009) 315;
  F.~Feruglio, C.~Hagedorn, Y.~Lin and L.~Merlo,
  arXiv:0808.0812 [hep-ph];
  R.~N.~Mohapatra and A.~Y.~Smirnov,
  Ann.\ Rev.\ Nucl.\ Part.\ Sci.\  {\bf 56} (2006) 569
  [arXiv:hep-ph/0603118].
  S.~F.~King,
  J.\ Phys.\ Conf.\ Ser.\  {\bf 136} (2008) 022038.
    
\end{thebibliography}
\end{document}